\documentclass[aps,epsf]{revtex4}
\usepackage{graphics}
\usepackage{graphicx}
\usepackage{epsfig}

\def\beq{\begin{equation}}
\def\eeq{\end{equation}}
\def\bea{\begin{eqnarray}}
\def\eea{\end{eqnarray}}

\newcommand{\D}{\displaystyle}

\begin{document}

\input epsf.tex

\title{Modeling a circular equatorial test-particle in a Kerr spacetime.}
\author{J\'er\^ome Carr\'e \& Edward K. Porter\\}
\vspace{1cm}
\affiliation{Fran\c{c}ois Arago Centre, APC, Universit\'e Paris Diderot,\\ CNRS/IN2P3, CEA/Irfu, Observatoire de Paris, Sorbonne Paris Cit\'e, \\10 rue A. Domon et L. Duquet, 75205 Paris Cedex 13, France}
\vspace{1cm}
\begin{abstract}
Extreme Mass Ratio Inspirals (EMRIs) are one of the main gravitational wave (GW) sources for a future space detector, such as eLISA/NGO, and third generation ground-based detectors, like the Einstein Telescope. These systems present an interest both in astrophysics and fundamental physics. In order to make a high precision determination of their physical parameters, we need very accurate theoretical waveform models or templates. In the case of a circular equatorial orbit, the key stumbling block to the creation of these templates is the flux function of the GW.  This function can be modeled either via very expensive numerical simulations, which then make the templates unusable for GW astronomy, or via some analytic approximation method such as a post-Newtonian approximation. This approximation is known to be asymptotically divergent and is only known up to 5.5PN order for the Schwarzschild case and to 4PN order for the Kerr case. A way to improve the convergence of the flux is to use re-summation methods. In this work we extend previous results using the Pad\'e and Chebyshev approximations, first by taking into account the absorption of the GWs by the central black hole which was neglected in previous studies, and secondly by using the information from the Schwarzschild and absorption terms to create a Kerr flux up to 5.5PN order. We found that these two additions both improve the convergence. We also demonstrate that the best re-summation method for improving the flux model is based on a flux function which we call the ``inverted Chebyshev approximation".
\end{abstract}

\maketitle


\section{Introduction.}

The Extreme Mass Ratio Inspirals (EMRIs) are one of the most important sources of gravitational waves (GWs) for future space detectors such as the eLISA/NGO mission~\cite{AABBB1,AABBB2}, and terrestrial detectors like the Einstein Telescope~\cite{ET1, ET2}. These systems are composed of a stellar mass compact object orbiting around a (super)massive black hole (BH) such that the mass ratio is on the order of $10^{-5}-10^{-7}$.  Their physical interest comes from the fact that the small object normally traces out approximately $10^5$ orbits from when it is first observed in the detector until it plunges into the central BH.  The test particle can then be thought of as something which maps the spacetime around the central BH, providing a very strong test of the theory of General Relativity.

Even if EMRIs are quite simple objects, their dynamics, which we need to understand for GW astrophysics, are very complicated.   As we do not have a solution for the case of generic orbits around the central BH, we concentrate on a test particle in a circular equatorial orbit.  Due to the fact that there is no plane precession in this case, we have one less constant of motion and the trajectories are completely described by the energy and angular momentum of the system.  One of the key aspects in  understanding the dynamics is the flux of GWs emitted by the system.  At present, we have an exact expression for the binding energy of the system, but for the flux we use a post-Newtonian (PN) approximation. This approximation is known it up to 4PN for the Kerr case~\cite{Poisson2,SSTT,TSTS} and up to 5.5PN in the Schwarzschild case~\cite{Poisson1,Cutetal1,TagNak,TagSas,TTS}.

While it was shown that the PN series breaks down for orbital separations of $r\leq10M$~\cite{Brady}, the validity of the PN flux was studied in the non-spinning and spinning cases respectively using an asymptotic analysis~\cite{YB,ZYB}. These studies found that for the Schwarzschild case the best PN expansion is at 3PN, but a conclusion could not be made for the Kerr case due to the lack of higher order terms in the series. The second result from these studies was that the PN approximation is more suitable for retrograde than for prograde orbits. This is most easily explained by the position of the last stable orbit (LSO).  When the dimensionless spin parameter, $q$, is positive, the LSO moves closer to the the central black hole as $q$ increases, reaching a maximum value of $r=M$ for $q=1$, and so, the system is more relativistic causing the PN approximation to break down sooner.   The contrary happens for $q<0$ as the position of the LSO moves outwards from the BH to a maximum value of $r=9M$ at a value of $q=-1$. And because the PN approximation is an expansion in the invariant PN velocity parameter, $x$, and assumes that $x<<c$, it makes sense that it is more convergent for less relativistic systems. Another interesting result is that the PN flux is so bad at small separations between the two bodies that it should not be used for $r/M$ smaller than a limit in the interval $r/M \in [6,54]$ depending on the spin of the central BH.  These results, along with previous efforts to improve the convergence of the PN series, are our motivation to resolve this problem. 

Given the starting information provided by the PN approximation, a way to improve the convergence of a series is to use re-summation methods.  The first proposed solution for the GW flux was to use a Pad\'e approximation, which replaces the PN series by a rational polynomial.  This was applied first to Schwarzschild orbits~\cite{DIS}, and was later extended to Kerr orbits~\cite{PS}. In both cases, they found that in general, the Pad\'e flux has a better and more monotonous convergence than the PN flux.   Even so, the Pad\'e solution was not complete due to the existence of poles in the approximation at certain PN orders, and also due to the fact that at some PN orders the new approximation was worse than the original PN series.  In previous articles it was demonstrated that one could create both sub and super-diagnonal Pad\'e approximations, depending on the number of terms in either the numerator or the denominator.  The problem of poles could be cured by moving to a sub-diagonal approximation should a pole occur in the super-diagonal approximation and vice versa.  However, due to the inherent form of the approximation, there is no cure for the cases where the approximation simply performs badly.  Furthermore, in the Kerr case, a variety of spins were investigated, but the flux was kept at the 4PN order, which we now know to be divergent.  The second solution which was proposed was to replace the divergent power series by a Chebyshev series, and was tested in the non-spinning test-mass and comparable mass cases~\cite{Porter1, Porter2}.  As a Chebyshev series is closely related to the minimax polynomial,  they were demonstrated to obtain a better convergence than using the PN or Pad\'e fluxes.  This improvement was due to two aspects : the first is that while power series have a radius of convergence that may not cover the entire domain of interest, Chebyshev series have an ellipse of convergence that spans the entire domain of interest.  Secondly, as we shall see later, the PN factorized flux contains logarithmic terms which begin at the 3PN order.  This part of the series is untreatable with Pad\'e approximation as it yields coefficients that are either null or infinite.  The Chebyshev approximation has no problems with the fact the the first non-zero term of the series is at 3PN order and we can thus define a Chebyshev series here as well.

In this work we extend these previous results, first, by taking into account the absorption of the GW by the central black hole. Secondly, in the spinning case, as the 4PN flux is divergent, we include the information that we have in the Schwarzschild case up 5.5PN in the Kerr flux function. Then, we search for the best factorized flux function, from the function proposed in Ref.~\cite{DIS} where the velocity at the LSO is used as a normalising velocity,  and of some of its variants, which include using the velocity at the photon ring as a normalising velocity or inverting the factorized flux function. Finally, we study the convergence of this more complete PN flux, and of the Pad\'e and Chebyshev flux models, for different orders of approximation and different spins.

We will see that the flux needs to be treated differently for different types of orbits : the non-spinning, retrograde and prograde cases. This separation stems from the differing relativistic nature involved as the test-particle approaches the central black hole and will influence our choice of normalizing velocity in the flux function. We will use the invariant velocity at the photon ring for the retrograde case, and the invariant velocity at the last stable orbit for the non-spinning and prograde cases. We will also see that in most situations, it is better to use an inverted factorized flux, introduced in Ref~\cite{PS}, rather than the original factorized flux, proposed by Ref~\cite{DIS}.   And finally we will see that for the Chebyshev re-summation, we will not necessarily use the highest known PN order, especially if that order less convergent than at lower orders.  

This work is organized as follow, in Section~\ref{PN_Flux} we will describe the PN flux. In Section~\ref{Resummation_method}, we will describe the Pad\'e and Chebyshev approximations, and we will show how to construct these respective fluxes. In Section~\ref{Results}, we will present our results for the different spins. Finally, in Section~\ref{Conclusion} we will provide our conclusion.


\section{Post-Newtonian GW flux.}
\label{PN_Flux}

Search algorithms for GW astronomy are based on the technique of matched filtering~\cite{Helstrom}, which cross-correlates the received signal with theoretical templates of known parameter values. In order to determinate as precisely as possible the physical parameters of the observed source, we need very accurate templates. As matched filtering is very sensitive to the GW phase, it is therefore imperative that we have a very accurate phase model. The GW phase can be written in the form
\beq
\phi(v) = \phi_0+2\int_v^{v_{lso}}v^3\frac{E'(v)}{F(v)}dv,
\eeq
where $\phi_0$ is a constant reference GW phase (which can be taken to be when the signal is first seen in the detector), $E(v)$ is the binding energy and $E'(v)=dE(v)/dv$. $F(v)$ is the flux of the GW, and $v$ is the local linear velocity given by $v\equiv \sqrt{M/r}$, where $r$ is the radial coordinate in Boyer-Lindquist coordinates and $M$ is the mass of the central black hole. In the case of equatorial and circular orbits, we have an exact analytical expression for the binding energy~\cite{BPT}
\beq
\label{eq:enegy}
E(v,q)=\eta\frac{1-2v^2+qv^3}{\sqrt{1-3v^2+2qv^3}},
\eeq
where $\eta=\mu M/(\mu+M)^2\sim\mu/M$ is the symmetric mass ratio, $\mu$ is the mass of the stellar mass object, and $q$ is the dimensionless spin parameter defined as $q\equiv a/M$.

From this expression we can determine three particular orbits. First, we can determinate the position of the last stable orbit (LSO) which corresponds to the condition $E'(v)=0$~\cite{BPT}
\begin{equation}
r_{lso}^{\pm}(q) =M\,\left[ 3 + z_{2}(q)\mp \sqrt{\left[ 3 - 
z_{1}(q)\right]\,\left[ 3 + z_{1}(q) + 2\,z_{2}(q) \right]}\right],
\end{equation}
where
\bea
z_{1}(q) & = & 1 + \left(1 - q^2\right)^{\frac{1}{3}}\left[\left(1 + q\right)^{\frac{1}{3}} + \left(1 - 
q \right)^{\frac{1}{3}}\right],\\ \nonumber \\
z_{2}(q) & = & \sqrt{3\,q^2 + z_{1}^{2}}.
\eea
The signs $+$ and $-$ correspond to prograde and the retrograde orbits respectively. Then we can obtain the last unstable orbit, also known as the photon ring, where the derivative of the binding energy, $E'(v)$, has a pole~\cite{BPT,Chand},
\begin{equation}
r_{pr}^{\pm}(q) = 2\,M \left[ 1 + \cos\left[ \frac{2}{3}\,\cos^{-1}\left(\mp q \right)\right]\right].
\end{equation} 
Finally, we can determine the position of the black hole horizon, where the energy is zero
\begin{equation}
r^{\pm}_{H} = M\left (1 + \sqrt{1 - q^{2}}\right ).
\end{equation}

The position of these three orbits varies with the sign and magnitude of $q$. In the Schwarzschild case the horizon, the photon ring and the last stable orbit are respectively at $2M$, $3M$ and $6M$. When $q$ is positive and increasing, these orbits move to smaller radius until the coincide at radius $r=M$ in the limiting case of a maximally spinning BH, i.e. $q=1$. When $q$ is negative and decreasing, the LSO moves away from the black hole, while the photon ring and the horizon again approach the central BH.

While we have a exact expression for the binding energy in the case of circular equatorial orbit, it is not the case for the GW flux function $F(v)$. Instead we use a PN approximation where we have a series approximation in powers of $v/c$. This approximation has been calculated up to 5.5PN order in the case of a Schwarzschild BH and up to 4PN in the case of a Kerr BH. The general form of the flux in both cases is given by the  expression~\cite{Poisson2,SSTT,TSTS,Poisson1,Cutetal1,TagNak,TagSas,TTS}
\beq
\label{eq:PN_flux}
F_{T_n}(x;q)=F_N(x)\left[ \sum_{k=0}^n a_k(q)x^k+\ln(x)\sum_{k=6}^nb_k(q)x^k\right],
\eeq
where $n=8$ and $11$ in the Kerr and Schwarzschild cases respectively, and $F_N(x)$ is the Newtonian flux given by
\beq
F_N(x)=\frac{32}{5}\eta^2x^{10}.
\eeq
Here, $x$ is the invariant PN velocity parameter observed at infinity, which is related to the orbital angular frequency $\Omega$ by $x=(M\Omega)^{1/3}$ (Note that this is a different use of notation to the comparable mass literature where we define $x=v^2 = (M\Omega)^{2/3}$). The relation between the invariant velocity $x$ and local linear velocity $v$ is given by~\cite{MSSTT}
\beq
x(v,q)=v(1-qv^3+q^2v^6)^{1/3},
\eeq
which reduces to $x=v$ in the Schwarzschild limit. The coefficients $a_k(q)$ and $b_k(q)$ appearing in the flux function are given in Appendix~\ref{PN_coefficients}. They are the sum of three contributions : a spinless term, a spin dependent term, and an absorption term~\cite{MSSTT}. The absorption term comes form the absorption of the GW by the central BH.

Until now, in the Kerr case, the two series in Eqn~(\ref{eq:PN_flux}) have been stopped at $n=8$ as we do not know the spin-dependent term for higher orders. However, this 4PN flux function is divergent.  In this work we propose an extension of this series up to $n=11$ where we use the spin-independent and the absorption terms for $n>8$.  While this flux function is still incomplete, due to the lack of knowledge of spin dependent coefficients at greater than 4PN order, we shall demonstrate that this flux function has a better convergence when compared with a numerical flux that the truncated 4PN Kerr flux function.


\section{Approximating the GW flux function.}
\label{Resummation_method}

The field of numerical analysis presents a number of ways of improving the convergence of a series.  Pad\'e approximation is a powerful method
due to the fact that it is the best way of representing a power series by means of a rational polynomial, i.e. 
\beq
\label{eq:Pade}
T_n(x)=\sum_{k=0}^na_kx^k \approx \frac{\displaystyle{\sum_{k=0}^Nb_kx^k}}{\displaystyle{\sum_{k=0}^Md_kx^k}}=P_M^N(x),
\eeq
with $n+1=N+M+1$ and the power series of the right hand side being identical to the original power series.  Theory suggests that if we think of the 
Pad\'e approximation as a matrix, the best convergence is achieved by staying as close to the diagonal as possible, i.e. keep the number of terms
in the numerator and denominator equal.  At odd orders of approximation, it is possible to allow either the number of terms in the numerator (super-
diagonal) or the denominator (sub-diagonal) increase by one term.  Thus, if we choose $M=N+\epsilon$, where $\epsilon=0$ or $1$, we have the 
diagonal and sub-diagonal approximations, respectively.  The advantage of the sub-diagnoal Pad\'e approximant is that we can write it in
the form of a continued fraction
\begin{equation}
P^N_{N+\epsilon} = \frac{\D \sum_{i=0}^N d_i x^i}{\D \sum_{i=0}^{N+\epsilon} f_i x^i}=\frac{c_{0}}{\D1 + 
                        \frac{c_{1}x}{\D1 + 
                         \frac{c_{2}x}{\D
                          \frac{\ddots}{{\D 1 + 
                                c_{n}x}}}}}.
\label{eq:confrac}
\end{equation}
In this form, when we go from the order $n$ to the order $n+1$, only the $(n+1)^{th}$ coefficient needs to be calculated~\cite{BO,CP}, all lower order coefficients remain
constant.  In contrast, with a super-diagonal approximation, each higher order of approximation requires a re-calculation
of all coefficients.  Pad\'e approximation is a very useful method for accelerating the convergence of a series as not only can it well represent zeros and 
poles, but it can provide new information regarding the underlying function.  The main problem with Pad\'e approximation is when it gets the approximation 
wrong.  As with any power series, the Pad\'e approximation is ultimately based on the Weierstrass theorem which only guarantees convergence with a high 
number of terms, it is impossible to know at any particular order if the Pad\'e approximation will perform well.

On the other hand, the best way of increasing the convergence of a power series by using a polynomial representation is to use the minimax polynomial, which
unfortunately is not that easy to find.  The minimax polynomial approximates the underlying function such that we are presented with an oscillating equal error
curve.  This approximation has the ability to allow the error to grow in parts of the domain where we already have high accuracy in order to improve the accuracy
in parts of the domain where the error begins to grow.  Closely related to the minimax polynomial is a polynomial based on Chebyshev polynomials of the first
kind which can be defined using the iterative relation
\beq
\label{eq:iterative_Cheb}
T_n(x)=2xT_{n-1}(x)-T_{n-2}(x),
\eeq
with
\beq
T_0(x)=1 \quad , \quad T_1(x)=x.
\eeq
The Chebyshev polynomials are orthogonal polynomials and are a special case of the Gegenbauer polynomials (see Ref~\cite{Porter1,MH} for a more in depth explanation).  They are defined in the interval $[-1,1]$, where at the $n^{th}$ order they have $n$ zeros and $n+1$ extrema such that $|T_n(x)|=1$.  As most problems are not limited to this interval and are defined in the interval $[a,b]$, so we define the shifted Chebyshev polynomials of the first kind, $T^*(x)=T(s)$, by
\beq
s=\frac{2x-(a+b)}{b-a}\,\,\,\,\,\,\,\,\,\,\,\,\,\,\forall\,\,\, x\in[a,b],s\in[-1,1].
\eeq
The shifted Chebyshev polynomials can be calculated using the following iterative expression
\beq
\label{eq:SCP}
T_n^*(x)=2\left(\frac{2x-(a+b)}{b-a}\right)T_{n-1}^*(x)-T_{n-2}^*(x),
\eeq
with
\beq
T_0^*(x)=1 \quad , \quad T_1^*(x)=\frac{2x-(a+b)}{b-a}.
\eeq
The Chebyshev approximation again tries to minimize the maximum error, albeit less successfully than the minimax polynomial, but sufficiently so to give a more
convergent series at lower orders of approximation than the truncated power series.  While a Pad\'e approximation can provide new information on an underlying
function, a Chebyshev approximation essentially re-packages the information that we have from the power series into a more useful form.  Another beauty of the
Chebyshev series is that each coefficient at higher order is smaller than the previous coefficient.  As the Chebyshev polynomials have a maximum magnitude of 
unity, this means that the truncation error associated with a Chebyshev series is essentially the value of the truncated coefficient.  Ultimately, the performance of
the Chebyshev series is dependent on the generating power series.  If a series at a particular order is particularly divergent, then the corresponding Chebyshev
series with perform better, but will still be divergent.  We will see that in this case, starting at a lower order of approximation in the power series can result in 
a more convergent Chebyshev approximation. 


\subsection{Pad\'e GW flux approximation.}
\label{Pade}

The first method that was introduced in order to try and improve the convergence of the flux function was based on a Pad\'e approximation~\cite{DIS}.  Due to a number of
issues with the PN approximation to the flux, a direct application of Pad\'e approximation is not possible.  The first problem with the PN flux are the logarithmic terms which appears at 3PN, which effect the convergence of the approximation.  In order to reduce this effect, it was shown that one could normalize the velocity parameter appearing in the logarithm and also factor out the logarithmic terms from the original series. This led to the creation of a new flux function
\beq
\label{eq:FF'}
F_{T_n}(x)=F_N(x)\left[ 1+\ln\left(\frac{x}{x_N}\right)\sum_{k=6}^nl_kx^k\right]\sum_{k=0}^n c_kx^k,
\eeq
where 
\beq
c_k = \left\{
\begin{array}{ll}
a_k  	& \quad \forall\,\,\, k<6\\
a_k + \ln(x_N)b_k & \quad \forall\,\,\, k\geq6
\end{array}\right. ,
\eeq
and
\beq
l_k = b_k - \sum_{i=1}^{k-6}c_il_{k-i}.
\eeq
In Ref~\cite{DIS}, the invariant velocity at the last stable orbit is used as the normalizing velocity (i.e. $x_N = x_{LSO}$).  This value was used as the intention was to model the inspiral of the test
particle only as far as the LSO.  As we approach this point the effect of the logarithmic terms are greatly diminished.  However, as there are no constraints in choosing the value of the normalizing velocity,  we will also investigate other possibilities.

A second issue is that the linear term of the PN flux is null, which creates a problem when we use the Pad\'e approximation.   A null term in a series can result in null and infinite Pad\'e coefficients when we write the expansion in continued fraction form.   Furthermore,  we know that the flux function has a pole at the photon ring, so in order that we use the full potential of the Pad\'e approximation it was required that the function exhibit this pole~\cite{NR,BO}. Therefore, by multiplying the flux by $\left(1-\frac{x}{x_{pr}}\right)$ we obtain the factorized flux
\beq
\label{eq:FF}
f_{T_N} = F_N(x)\left[ 1+\ln\left(\frac{x}{x_N}\right)\sum_{k=6}^nl_kx^k\right]\sum_{k=0}^n f_kx^k,
\eeq
where
\bea
f_0 & = & c_0,\\
f_k & = & c_k-\frac{c_{k-1}}{x_{pr}},
\eea
and where $x_{pr}$ is the invariant velocity at the photon ring.

There exits another factorized flux introduced in Ref~\cite{PS}, that we will call the inverted factorized flux, which is obtained by analytically inverting the previous equation, which gives,
\beq
\label{eq:IFF}
f_{IT_n}(x)=\frac{1}{F_N(x)}\left[ 1-\ln\left(\frac{x}{x_{lso}}\right)\sum_{k=6}^nl_kx^k\right]\sum_{k=0}^n d_kx^k,
\eeq
where
\beq
\sum_{k=0}^n d_kx^k = \left(\sum_{k=0}^n f_kx^k\right)^{-1},
\eeq
and the $d_k$ coefficients are calculated using
\bea
d_0 & = & \frac{1}{f_0}, \\
d_k & = & -\frac{1}{f_0}\sum_{j=1}^kf_jd_{k-j}.
\eea
As the sum over the $l_k$'s in Eqn~(\ref{eq:FF}) and Eqn~(\ref{eq:IFF}) only starts at $k=6$, we cannot apply the Pad\'e approximation as it would generate null, undefined and infinite coefficients in the Pad\'e approximation. So the Pad\'e approximation is only applied to the non-logarithmic sum in the above equations, and we hence obtain the Pad\'e flux~\cite{DIS}
\beq
\label{eq:PF}
F_{P_n}(x)=F_N(x)\left(1-\frac{x}{x_{pr}}\right)^{-1}\left[ 1+\ln\left(\frac{x}{x_N}\right)\sum_{k=6}^nl_kx^k\right]P_{N+\epsilon}^N\left[\sum_{k=0}^n f_kx^k\right],
\eeq
and the inverted Pad\'e flux~\cite{PS}
\beq
\label{eq:IFP}
F_{IP_n}(x)=F_N(x)\left[\left(1-\frac{x}{x_{pr}}\right)\left[ 1-\ln\left(\frac{x}{x_N}\right)\sum_{k=6}^nl_kx^k\right]P_{N+\epsilon}^N\left[\sum_{k=0}^n d_kx^k\right]\right]^{-1}.
\eeq
The Pad\'e approximant was studied in the non-spinning~\cite{DIS} and spinning~\cite{PS} cases, and it was shown that in both case the Pad\'e flux better modeled a numerical flux than the PN flux.  However it was also seen in both cases that the Pad\'e approximation can suffer from singularities at different PN orders.  In principle we can recover
from this by moving to a super-diagonal approximant.  Furthermore, in some cases the Pad\'e approximation was a worse fit than the original PN series.  This was due to 
the fact that the factorized flux is highly divergent at certain PN orders and performed much worse than the PN series.  While the Pad\'e approximation was able to improve the convergence of the factorized flux, it still remained less convergent than the PN flux.  This also turned out to be the case in the Kerr study.  The convergence of the flux was studied for different spin values of the central BH, but only at the highest Kerr order of 4PN.  However, we now know that this order is more divergent than lower orders and thus the study may not be truly reflective of just how well we can model the Kerr flux function.


\subsection{Chebyshev flux approximation.}

To model the Chebyshev flux,  we defined the interval between the initial and LSO velocities $[x_{ini},x_{lso}]$, where we now further define two variables
\bea
\chi = x_{lso} + x_{ini}, \\
\xi = x_{lso} - x_{ini}.
\eea
From Eqn~(\ref{eq:SCP}), we have the new iterative relation
\beq
T_n^*(x)=2\left(\frac{2x-\chi}{\xi}\right)T_{n-1}^*(x)-T_{n-2}^*(x),
\eeq
with
\beq
T_0^*(x)=1 \quad , \quad T_1^*(x)=\frac{2x-\chi}{\xi}.
\eeq
For our study we took $x_{ini}=0$, however we could just as easily take the velocity when the system is first seen in the detector.

Contrary to the Pad\'e approximation, the Chebyshev re-summation is not effected by null coefficients and can be applied to both sums appearing in the factorized flux function.  We therefore define the Chebyshev flux function as~\cite{Porter1}
\beq
\label{eq:FC}
F_{C_n}(x)=F_N(x)\left(1-\frac{x}{x_{N}}\right)^{-1}\left[ 1+\ln\left(\frac{x}{x_{N}}\right)\sum_{k=0}^n\mathcal{L}_kT^*_k(x)\right]\sum_{k=0}^n \mathcal{F}_kT^*_k(x),
\eeq
with the Chebyshev series
\beq
\sum_{k=0}^n\mathcal{L}_k(l_k)T^*_k(x) = \sum_{k=6}^nl_kx^k
\eeq
and
\beq
\sum_{k=0}^n \mathcal{F}_k(f_k)T^*_k(x) = \sum_{k=0}^n f_kx^k.
\eeq
As the inverted Pad\'e flux was more convergent when compared with the numerical flux, we thus similarly define an inverted Chebyshev flux function
\beq
\label{eq:IFC}
F_{IC_n}(x)=F_N(x)\left[\left(1-\frac{x}{x_{pole}}\right)\left[ 1-\ln\left(\frac{x}{x_{N}}\right)\sum_{k=0}^n\mathcal{L}_kT^*_k(x)\right]\sum_{k=0}^n\mathcal{D}_kT^*_k(x)\right]^{-1},
\eeq
with 
\beq
\sum_{k=0}^n\mathcal{D}_k(d_k)T^*_k(x) = \sum_{k=0}^nd_kx^k.
\eeq

The Chebyshev re-summation was studied in the Schwarzschild case~\cite{Porter1}, and in this case it has already shown a more stable behavior which assured the minimization of the error between the approximated and numerical fluxes when the order of approximation increased.  In general the Chebyshev approximation provided better accuracy at lower orders than either the PN or Pad\'e fluxes. The only approximation at which it did not perform the best was at the  5.5PN order where it was better than the PN flux but worse than the Pad\'e flux, which for unknown reasons is particularly good.  We believe that as the Pad\'e approximation had a highly oscillatory nature of performance, it just happened to match the numerical flux closely at this PN order.  We do not believe that this was a sign of convergence of the Pad\'e approximation as the 6PN approximation (if we had it) could again have been worse than the PN approximation.  While it is not truly reflective as a metric due to the fact that a bad approximation could luckily coincide with the numerical flux, the error at the LSO was also investigated for each approximation.  While the PN and Pad\'e fluxes oscillated wildly in their performance, the Chebyshev approximation essentially reached the same constant error value at all orders of approximation which was at least a factor of two better, and sometimes a factor of 13 better, than either the PN or Pad\'e flux models.


\section{Results.}
\label{Results}

We know that depending on the value of the spin, the position of the LSO is different. For prograde orbits, the test-particle approaches closer to the central black and the system is more relativistic, whereas for retrograde orbits the small body falls into the black hole earlier, and the system is less relativistic. Based on this physical consideration, we decided to organize our study into three parts~: the non-spinning case, the retrograde case and the prograde case. In each case, we found the best factorized flux for each re-summation technique, and using these best-choice fluxes we compared the two re-summed fluxes  with the PN approximation at different orders and for different spins.


\subsection{The Schwarzschild Case.}

\begin{figure}[t]
\begin{center}
\epsfig{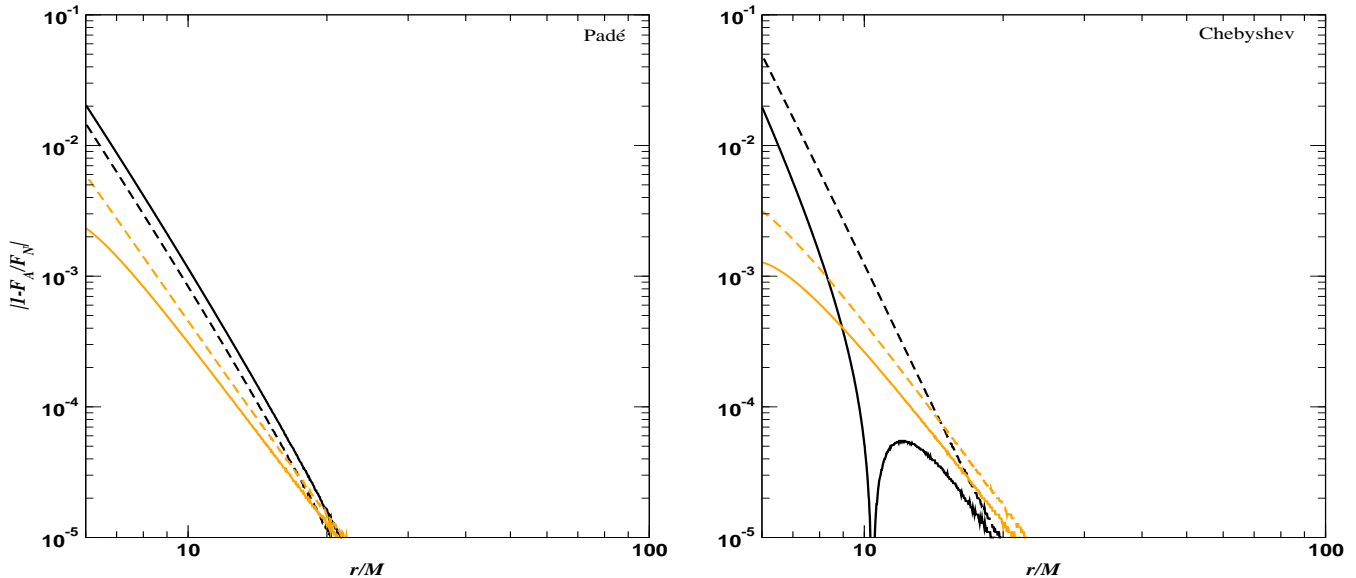}
\end{center}
\caption{Relative error between the approximated fluxes, $F_A$, and the numerical flux, $F_N$, in the non-spinning case at 5.5PN for the  Pad\'e (left panel) and Chebyshev (right panel) flux functions.  The dark (black) lines correspond to the factorized flux and the light (orange) line correspond to the inverted factorized flux. The solid lines represent a normalization of the logarithmic term using $x_{lso}$ and the dashed lines represent a normalization using $x_{pr}$. The best choice in both cases is the inverted flux using $x_{lso}$ for the normalization.}
\label{fig:FF_Schwarzschild}
\end{figure}

Our first study concerns the non-spinning case, and before working on any resummation methods we first need to find the best factorized flux which will form the basis of these methods. We saw in Section~\ref{Pade} that two aspects of the factorized flux can been tuned~: the choice of the normalization in the logarithmic term (we tried $x_{lso}$ and $x_{pr}$) and the choice to use an inverted or an non-inverted factorized flux. We tried these four cases at 5.5PN and we applied on each of them the Pad\'e and the Chebyshev re-summations, the results are shown in Fig~\ref{fig:FF_Schwarzschild}. The first thing that we see is that for both Pad\'e and Chebyshev fluxes, and for the two normalizations, the inverted flux models have a smaller error than the non-inverted flux.  Depending on which normalization factor is used, the improvement ranges between a factor of two and an order of magnitude. This validates the proposition made in Ref~\cite{PS} to use the inverted flux. We also see that for both the Pad\'e and Chebyshev fluxes,  using $x_{lso}$ for the normalization, which was the choice made in~\cite{DIS}, produces a smaller maximum error. Consequently, the best flux, in this sense, is the inverted flux using $x_{lso}$ for the normalization.  While the results presented here are for the 5.5PN case, we have confirmed a similar behaviour at lower orders of approximation.

\begin{figure}[t]
\begin{center}
\epsfig{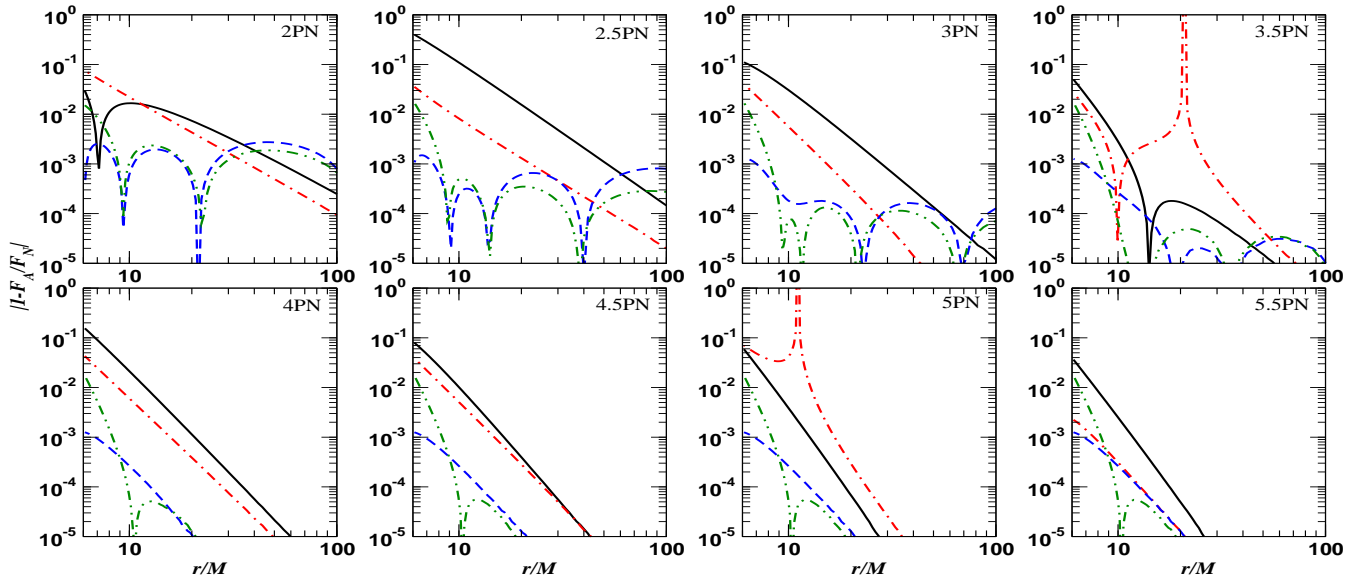}
\end{center}
\caption{Relative error between the approximated flux, $F_A$, and the numerical flux, $F_N$, in the non-spinning case, for PN flux (black solid curve), the inverted Pad\'e flux (red dash-dotted curve), the Chebyshev flux (green double-dot-dashed curve) and the inverted Chebyshev flux (blue dashed curve), for all PN orders from 2PN to 5.5PN.}
\label{fig:Schwarzschild}
\end{figure}

Using this factorized flux we compare the PN (black solid curves), Pad\'e (red dot-dashed curves),  Chebyshev (green double-dot-dashed curves) and inverted Chebyshev (blue dashed curves) fluxes with a numerical flux. This comparison ranges from 2PN to 5.5PN and is shown in Fig~\ref{fig:Schwarzschild}. The first thing that we see is that, except at the 2PN order where the PN flux has a fortunate coincidence with the numerical flux and the orders where the Pad\'e flux has a singularity,  the three re-summations are always more convergent than the PN flux.   If we compare the performance of each flux model as we approach the last stable orbit (and again keeping in mind that this is not on its own the best metric of convergence) we can see that the PN approximation returns a relative error $\epsilon=|1-F_A/F_N|$ (where $F_A$ is the approximated flux, and $F_N$ is the numerical flux) of $0.03 \leq \epsilon \leq 0.4$ at various orders of approximation.  The Pad\'e approximation, when not suffering from a singularity, has relative errors of between $3\times10^{-2}\leq \epsilon \leq 8\times10^{-2}$, except at the 5.5PN order where it has an error of $1.7\times10^{-3}$.  The Chebyshev approximation, as expected, tries to minimize the maximum error and produce an oscillating error curve.  As was shown in Ref~\cite{Porter1}, the Chebyshev approximation converges to an almost constant error value of $1.8\times10^{-2}$ at the LSO.  The inverted Chebyshev approximation performs fantastically in this case achieving a constant error value of approximately $\epsilon=10^{-3}$ at all orders of approximation.  Therefore, we can see from the figure that the inverted Chebyshev flux at 2PN order is as accurate as the 5.5PN Pad\'e approximation (that is difficult to explain) and at least an order of magnitude better than all other models at each order of approximation.

At wider separations, the Chebyshev approximation allows the error to float a little, but in this domain, we do not require the same level of accuracy as we do closer to the central BH.

\subsection{Retrograde Orbits.}

\begin{figure}[t]
\begin{center}
\epsfig{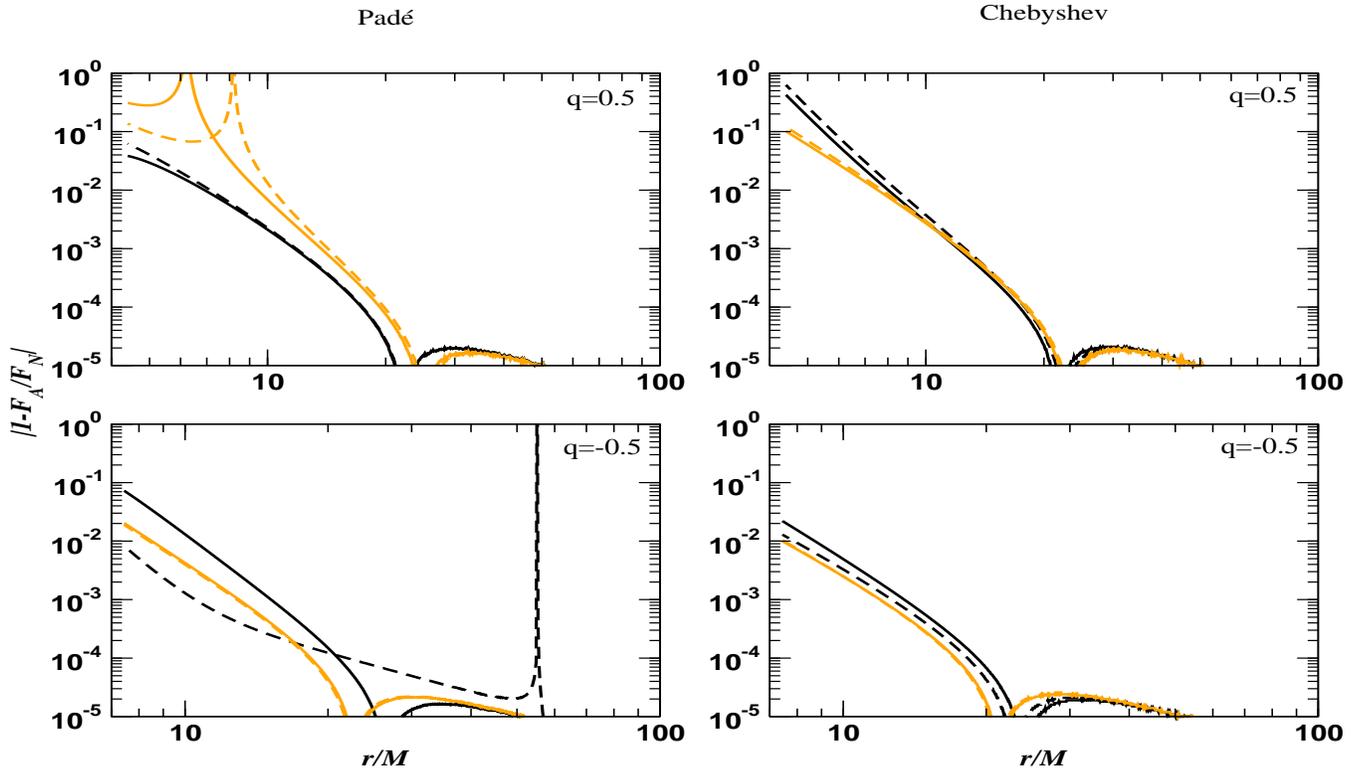}
\end{center}
\caption{Relative error between the approximated flux, $F_A$, and the numerical flux, $F_N$, in the spinning case at 5.5PN for the Pad\'e (left panels) and Chebyshev (right panels) flux functions for spins of $q=0.50$ (top panels) and $q=-0.50$ (bottom panels). The dark (black) lines correspond to the factorized flux and the light (orange) lines correspond to the inverted factorized flux. The solid lines represent a normalization of the logarithmic term useing $x_{lso}$, while the dashed lines represent a normalization by $x_{pr}$. For the prograde case the best choices are the the non-inverted flux for the Pad\'e approximation and the inverted flux for the Chebyshev approximation, with both  fluxes using $x_{lso}$ for the normalization. For the retrograde case, the best choice is the inverted flux using $x_{pr}$ for the normalization, for both Pad\'e and Chebyshev approximations.}
\label{fig:FF_Kerr}
\end{figure}

As in the Schwarzschild case, we start by searching for the best factorized flux for each re-summation method. We demonstrate in the bottom two cells of Fig~\ref{fig:FF_Kerr}, the comparison between the inverted/non-inverted fluxes, using different normalization constants, at 5.5PN for a spin $q=-0.5$. The result from this case is representative of the results for all other retrograde spins.  As in the Schwarzschild case, we see that once again for both the Pad\'e and Chebyshev flux models,  the inverted flux is the more convergent model. One could argue in the Pad\'e case that the standard flux looks to be the best choice as it has a smaller error as we approach the LSO, but it is not the case due to the presence of a singularity in the flux. This pole is also present at all other spins and other differing orders of approximation, which is why we reject using this flux. We also can see that using $x_{pr}$ is a marginally better choice than using $x_{lso}$ as a normalizing velocity.  However, the results are so close that one could continue to use the LSO value as a normalizing factor. 

\begin{figure}[t]
\begin{center}
\epsfig{file=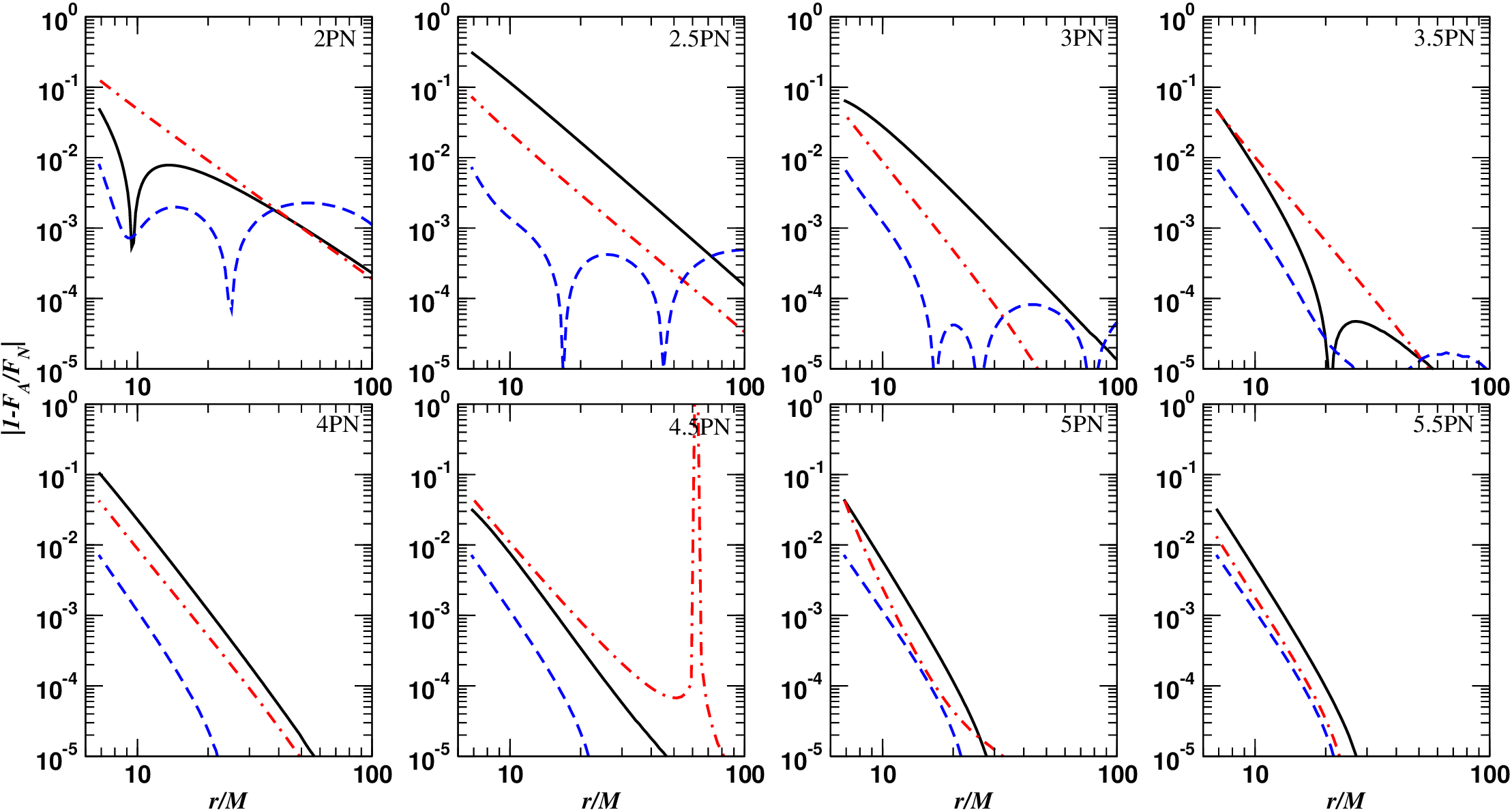, width=7in, height=3.in}
\end{center}
\caption{Relative error between the approximated flux, $F_A$, and the numerical flux, $F_N$, with $q=-0.25$, for the PN  (black solid curve),  inverted Pad\'e  (red dash-dotted curve) and inverted Chebyshev (blue dashed curve) flux models at all PN orders from 2PN to 5.5PN.}
\label{fig:0.25cr}
\end{figure}

\begin{figure}[t]
\begin{center}
\epsfig{file=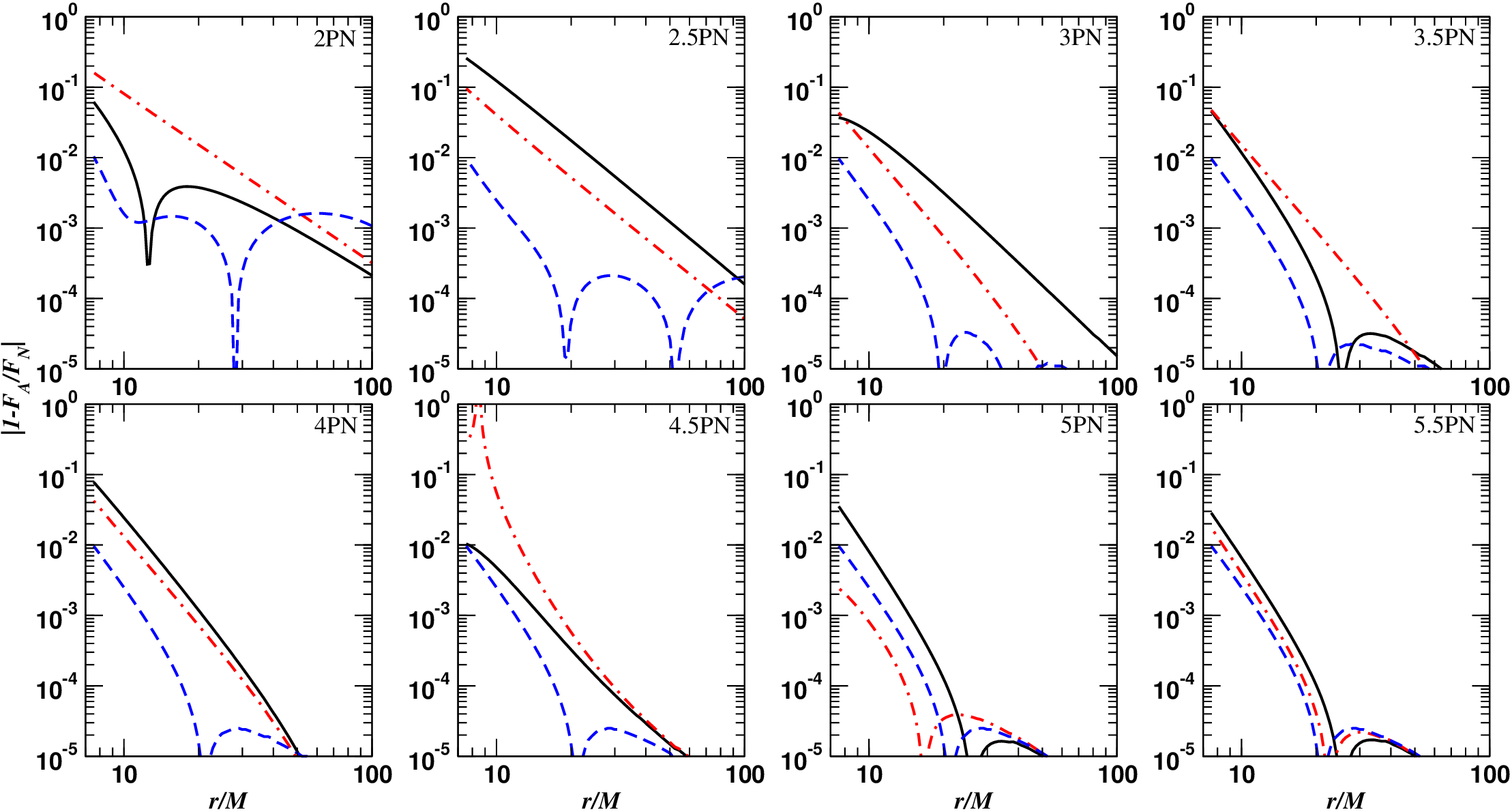, width=7in, height=3.in}
\end{center}
\caption{Relative error between the approximated flux, $F_A$, and the numerical flux, $F_N$, with $q=-0.50$, for the PN (black solid curve),  inverted Pad\'e (red dash-dotted curve) and inverted Chebyshev (blue dashed curve) flux models at all PN orders from 2PN to 5.5PN.}
\label{fig:0.50cr}
\end{figure}

\begin{figure}[t]
\begin{center}
\epsfig{file=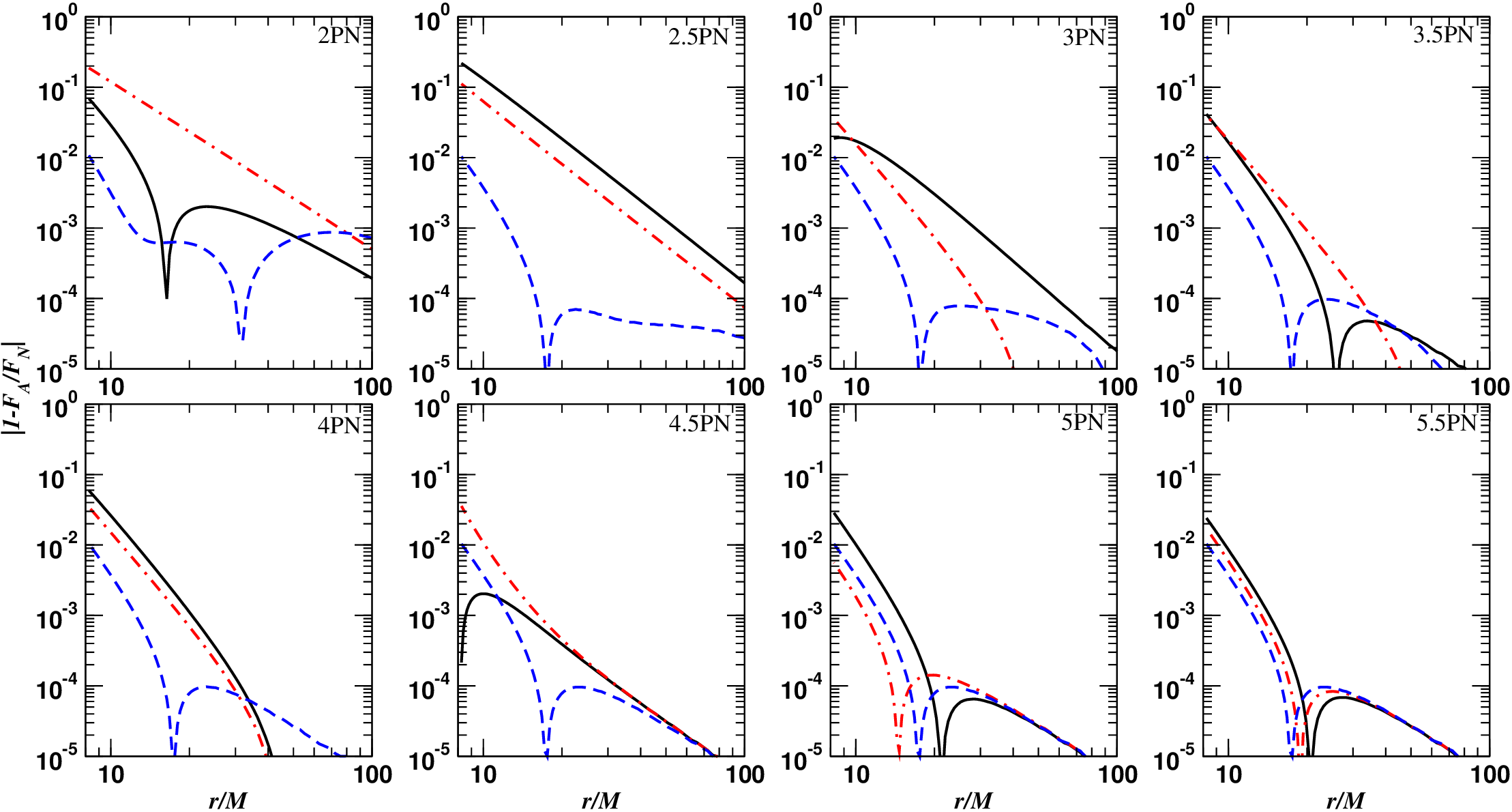, width=7in, height=3.in}
\end{center}
\caption{Relative error between the approximated flux, $F_A$, and the numerical flux, $F_N$, with $q=-0.75$, for the PN (black solid curve),  inverted Pad\'e (red dash-dotted curve) and inverted Chebyshev (blue dashed curve) flux models at all PN orders from 2PN to 5.5PN.}
\label{fig:0.75cr}
\end{figure}

\begin{figure}[t]
\begin{center}
\epsfig{file=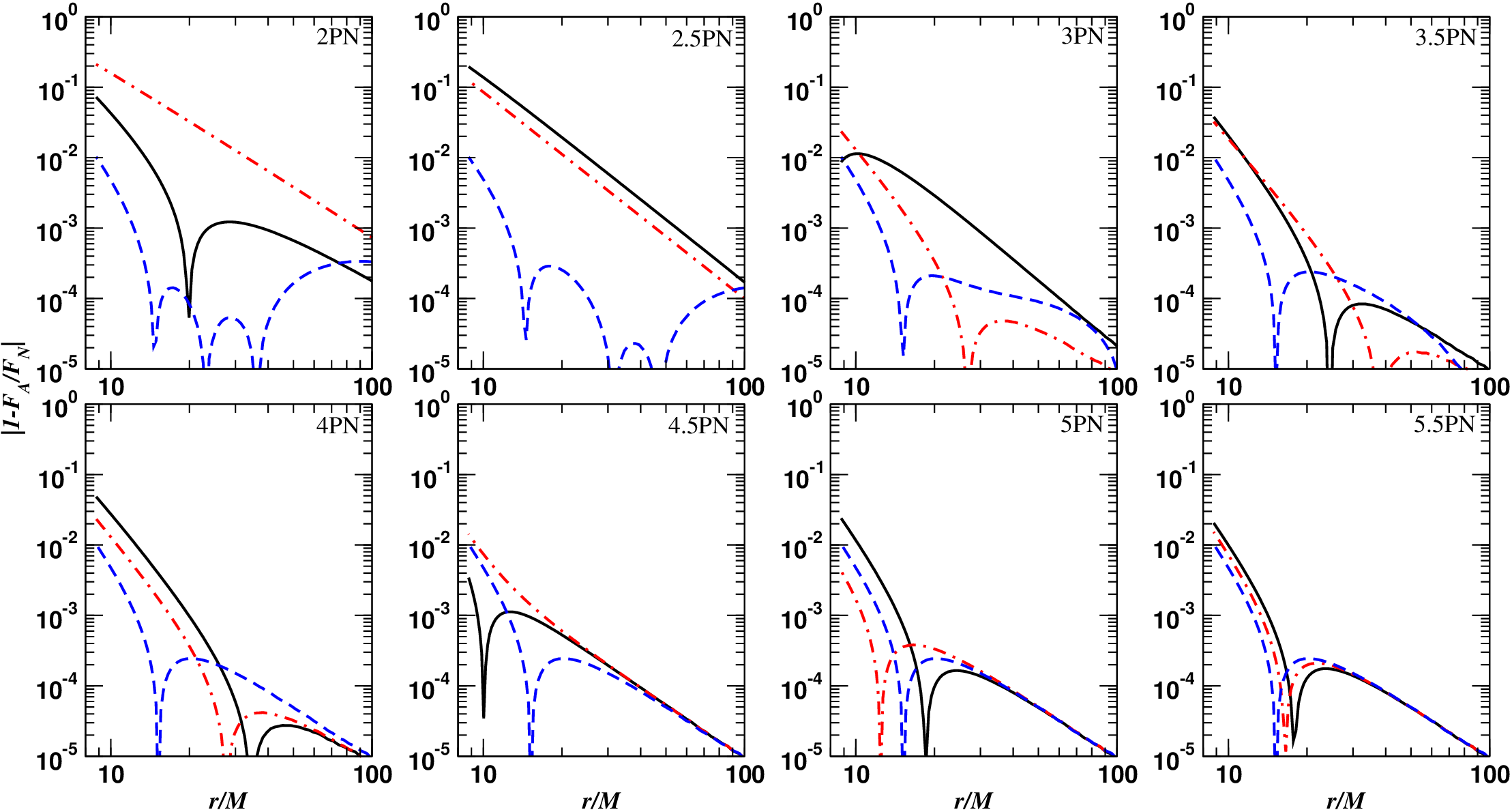, width=7in, height=3.in}
\end{center}
\caption{Relative error between the approximated flux, $F_A$, and the numerical flux, $F_N$, with $q=-0.95$, for the PN (black solid curve),  inverted Pad\'e (red dash-dotted curve) and inverted Chebyshev (blue dashed curve) flux models at all PN orders from 2PN to 5.5PN.}
\label{fig:0.95cr}
\end{figure}

Now that we have a model for the optimal factorized flux, we now compare the PN flux (black solid curves), the inverted Pad\'e (red dot-dashed curves) and the inverted Chebyshev (blue dashed curves), from 2PN to 5.5PN for the four retrograde spin values of $q=-0.25,-0.50,-0.75,-0.95$, which are shown in Fig~\ref{fig:0.25cr}-\ref{fig:0.95cr}. The first thing that we notice is that the addition of the Schwarzschild and absorption terms to the PN flux to 5.5PN order provides a more convergent PN flux model than the standard PN Kerr model at 4PN.  This result has the knock-on effect that the ensuing 5.5PN factorized flux is better performing than the corresponding factorized flux at 4PN order, thus ensuring that the re-summed fluxes starting at 5.5PN order are more convergent than previously studied.

If we first investigate the performance of the PN flux, we see that this approximation has a slightly improved convergence at 5.5PN order as we move from a spin of -0.25 to a spin of -0.95.  This is due to the fact that the higher retrograde spins being less relativistic and the test-particle plunging further out from the central BH.  In general, at $q=-0.25$ the error at the LSO is between $3\times10^{-2}\leq \epsilon \leq 0.3$, improving to between $10^{-2}\leq \epsilon \leq 0.2$ at a spin of $q=-0.95$.  While the performance of the PN flux illustrates the oscillatory convergence that we would expect, we should point out that as we move towards a maximally counter-rotating spin, the 4.5PN order PN approximation performs better and better with an errors of $3\times10^{-2},10^{-2}, 2\times10^{-4}$ and $3\times10^{-3}$ at increasing orders of spin.  This mimics the unexplainable behaviour we observed in the Schwarzschild case for the Pad\'e approximation at 5.5PN order, and in this case the PN flux just happens to have a fortunate coincidence with the numerical flux as we approach the LSO.

The inverted Pad\'e flux does not perform particularly well in the case of retrograde motion.  At all spin values, when not suffering from singularities, there are levels of approximation where the Pad\'e approximation is inferior to the original PN flux function.  This is due to the fact that the factorized flux is divergent at these orders and performs much worse than the original PN flux function.  In the cases where it does provide an improvement, the overall gain is a factor approximately 1.5 over the PN approximation with errors of between $2\times10^{-2}\leq \epsilon \leq 0.2$ and $10^{-2}\leq \epsilon \leq 0.1$ at  spins of -0.25 and -0.95 respectively.

The inverted Chebyshev approximation performs consistently well at all spin values, with relative errors of $\epsilon=7\times10^{-3}$ at $q=-0.25$ and $\epsilon=10^{-2}$ at a spin of $q=-0.95$.  Regardless of the spin value, the inverted Chebyshev flux is approximately an order of magnitude better at low orders of approximation and at least a factor of two better at higher orders.  More importantly is the fact that this approximation is almost always better than both the PN and inverted Pad\'e fluxes, and essentially converges to the same order of magnitude error at the LSO for all retrograde spin values.  In the cases where it is less convergent than the other approximations, we again highlight the fact at times the PN approximation has a fortunate coincidence with the numerical flux as we approach the LSO, and the Pad\'e approximation at 5PN order seems to be very good for high retrograde spin values.  However, neither of these facts can be considered as global trends, so it is clear that for retrograde orbits, the inverted Chebyshev flux provides a more convergent flux model than either of the other two methods.  Furthermore, the inverted Chebyshev flux function at 2.5 or 3PN order appears to have a sufficiently small error at all values of separation to be sufficient as a basis for a template.


\subsection{Prograde Orbits.}

We finish our study by looking at the prograde case. As previously, we first search for the best factorized flux. In the top two panels of Fig~\ref{fig:FF_Kerr} we plot the comparison between the Pad\'e and Chebyshev fluxes at 5.5PN order for $q=0.5$. We see that in both cases the normalization using $x_{lso}$ is marginally better than a normalization using $x_{pr}$. In the Pad\'e case the best choice of flux is the standard Pad\'e flux due to the presence of the poles in both of the inverted flux models. On further investigation, these poles appear for all values of prograde spin, so we choose the non-inverted factorized flux as the basis for the Pad\'e approximation. On the other hand, for the Chebyshev flux, there is not much difference between the four flux models until we approach the LSO.  In this case that standard Chebyshev flux, with either normalization, begins to diverge.  We will therefore use the inverted Chebyshev flux with $x_{lso}$ as a normalizing velocity as our preferred model.

\begin{figure}[t]
\begin{center}
\epsfig{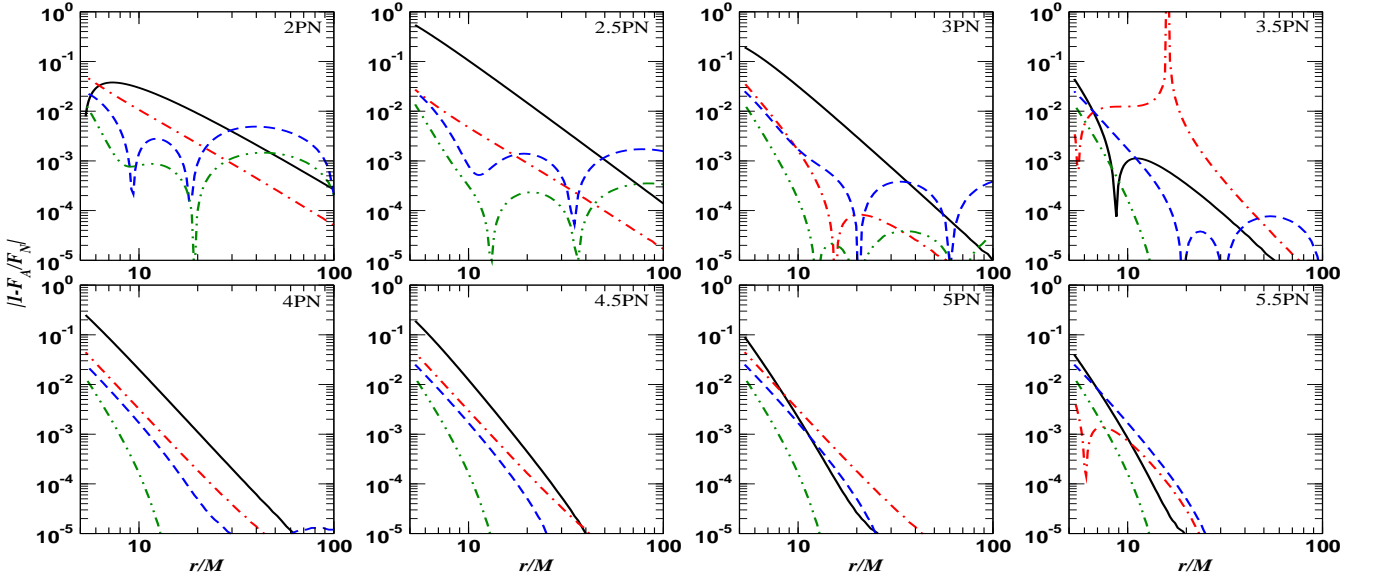}
\end{center}
\caption{Relative error between the approximated flux, $F_A$, and the numerical flux, $F_N$, with $q=0.25$, for PN (black solid curve), Pad\'e (red dash-dotted curve),  inverted 5.5PN Chebyshev (blue dashed curve) and inverted 5PN Chebyshev (green double-dot-dashed curve) flux models at all PN orders from 2PN to 5.5PN. In the last panel, the green double-dot-dashed curve is again the inverted 5PN Chebyshev flux as a comparison.}
\label{fig:0.25}
\end{figure}

\begin{figure}[t]
\begin{center}
\epsfig{file=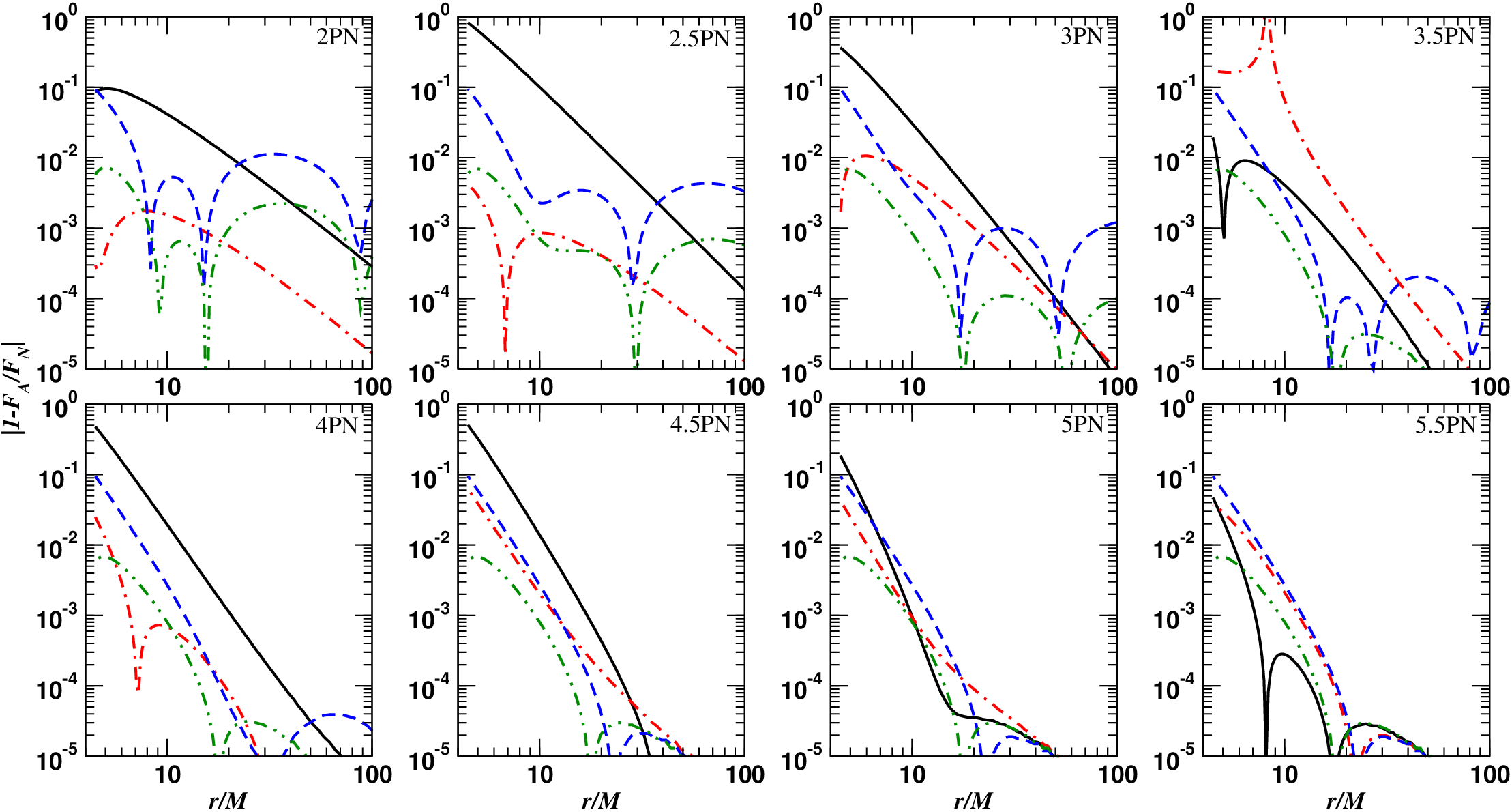, width=7in, height=3.in}
\end{center}
\caption{Relative error between the approximated flux, $F_A$, and the numerical flux, $F_N$, with $q=0.50$, for PN (black solid curve), Pad\'e (red dash-dotted curve),  inverted 5.5PN Chebyshev (blue dashed curve) and inverted 5PN Chebyshev (green double-dot-dashed curve) flux models at all PN orders from 2PN to 5.5PN. In the last panel, the green double-dot-dashed curve is again the inverted 5PN Chebyshev flux as a comparison.}
\label{fig:0.50}
\end{figure}

\begin{figure}[t]
\begin{center}
\epsfig{file=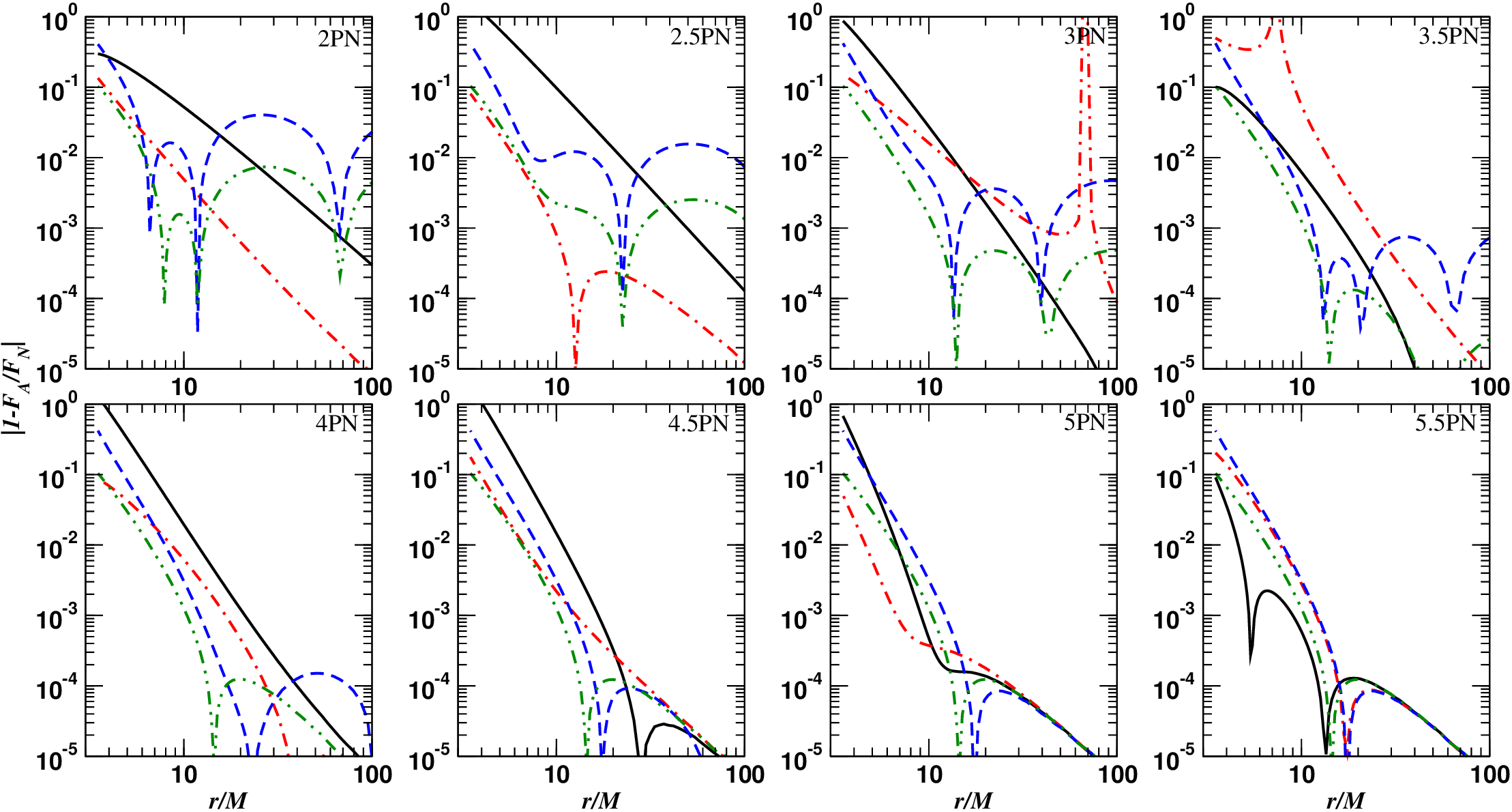, width=7in, height=3.in}
\end{center}
\caption{Relative error between the approximated flux, $F_A$, and the numerical flux, $F_N$, with $q=0.75$, for PN (black solid curve), Pad\'e (red dash-dotted curve),  inverted 5.5PN Chebyshev (blue dashed curve) and inverted 5PN Chebyshev (green double-dot-dashed curve) flux models at all PN orders from 2PN to 5.5PN. In the last panel, the green double-dot-dashed curve is again the inverted 5PN Chebyshev flux as a comparison.}
\label{fig:0.75}
\end{figure}

\begin{figure}[t]
\begin{center}
\epsfig{file=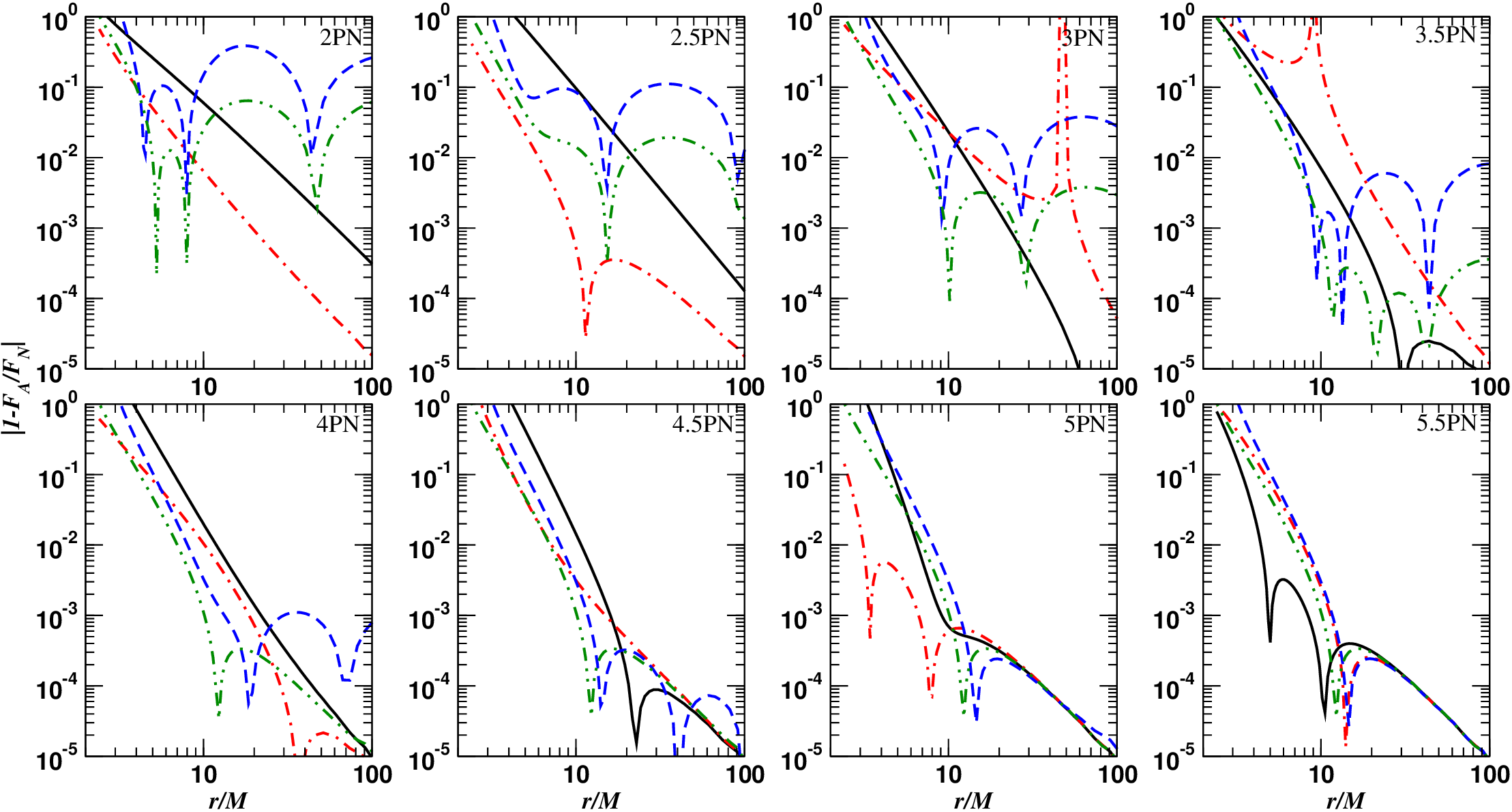, width=7in, height=3.in}
\end{center}
\caption{Relative error between the approximated flux, $F_A$, and the numerical flux, $F_N$, with $q=0.95$, for PN (black solid curve), Pad\'e (red dash-dotted curve),  inverted 5.5PN Chebyshev (blue dashed curve) and inverted 5PN Chebyshev (green double-dot-dashed curve) flux models at all PN orders from 2PN to 5.5PN. In the last panel, the green double-dot-dashed curve is again the inverted 5PN Chebyshev flux as a comparison.}
\label{fig:0.95}
\end{figure}

As for the retrograde case we compare the PN (black solid curves), Pad\'e (red dot-dashed curves) and the inverted Chebyshev fluxes (blue dashed curves), from 2PN to 5.5PN and for $q=0.25,0.50,0.75,0.95$, in Fig~\ref{fig:0.25}-\ref{fig:0.95}. First of all, we can see that, contrary to the retrograde case, when the spin increases the convergence of the flux decreases. This is explained by the fact that when the spin increases, the test-particle is now orbiting much closer to the central black hole and the system becomes much more relativistic. In this case the PN flux breaks down faster.  Before discussing the results in full, we should mention here that during our investigation, we noticed that the inverted Chebyshev flux was not performing as well as we had expected.  When delving deeper into this issue, we discovered that the factorized flux function at 5.5PN order behaved much worse than the original PN approximation.  The Chebyshev approximation, while being able to improve on this somewhat, was unable to recover from the high divergence of the factorized flux.  We then studied the behaviour of the factorized flux at different PN orders and discovered that the factorized flux at 5PN order was better performing than the 5.5PN PN flux.  As the performance of the Chebyshev approximation is highly dependent on the starting information, we decided to construct the inverted Chebyshev flux using the factorized flux at 5PN order.  In the following figures, a new curve represented this truncated inverted Chebyshev flux is represented (green double-dot-dashed curves).

The case of prograde orbits is sufficiently difficult to describe that we divide it into two regimes.    We noticed that even up to a spin of $q=0.5$ the orbits are only slightly more relativistic than the Schwarzschild case, while from spins of $q=0.75$ onwards, the orbits become truly relativistic to the point that the PN flux begins to break down sooner and sooner.  However, one global observation is that, even in the prograde case, the PN flux at 4.5 to 5.5PN order is more convergent than the standard Kerr 4PN flux.  This again justifies the inclusion of the Schwarzschild and absorption terms at higher PN orders.

Firstly, we will discuss the results at spins of 0.25 and 0.5.  In both these cases, the PN flux performs as expected with fractional errors of between $8\times10^{-3}\leq \epsilon \leq 0.5$ and $2\times10^{-2}\leq \epsilon \leq 0.9$ respectively.  In the lower spin case the error of $\epsilon = 8\times10^{-3}$ is due again to a lucky coincidence with the numerical flux as we approach the LSO.  In general the mean error is on the order of $\epsilon = 0.1$.  In the $q=0.5$ case, things begin to become a little clearer and we start to oscillate wildly between highly convergent and highly divergent.  While the Pad\'e flux presents a large improvement on the PN flux, in terms of stability, we still see some of the wild fluctuations associated with the PN flux, with relative errors of between $4\times10^{-3}\leq \epsilon \leq 5\times10^{-2}$ and $3\times10^{-4}\leq \epsilon \leq 6\times10^{-2}$.  Here as in the PN case, the extremely convergent values are due coincidences with the numerical fluxes as we approach the LSO.  In general the errors are of the order of $\epsilon = 5\times10^{-2}$ and $\epsilon = 3\times 10^{-2}$ respectively for both spin values.   At this level of spin, the Pad\'e flux suffers from singularities at the 3.5PN order which rule out using these orders for template creation.  The inverted Chebyshev flux starting from the divergent factorized flux at 5.5PN order is almost always better than the PN flux at a spin of $q=0.25$ achieving an almost constant error of $\epsilon = 2\times10^{-2}$ at the LSO.  For this low spin, it in general also outperforms the Pad\'e approximation.  However for a spin of $q=0.5$, while still outperforming the PN flux function, the 5.5PN inverted Chebyshev flux is constantly worse than the Pad\'e approximation, achieving a maximum spin error of only $\epsilon=0.1$.  This is also reflected at higher spin values and demonstrates the increasing divergence of the 5.5PN factorized flux.  However, if we now concentrate on the 5PN inverted Chebyshev flux, we can see that it approaches an almost constant error value of $\epsilon=10^{-2}$ for a spin of $q=0.25$ and  $\epsilon=5\times10^{-3}$ for a spin of $q=0.5$.  For both spin values, it is consistently better than all other approximation models.  While in one or two cases it is beaten by a fortunate PN or Pad\'e flux model, it clearly provides a more stable model, even at higher values of $r/M$.

We next move onto the more relativistic spins of $q=0.75$ and $q=0.95$.   For these two spin values, there is not much point in discussing the relative errors for the various approximations as none of them can cope with the high relativistic nature of the orbits.  In this case, a more indicative metric may be the separation at which the relative error reaches $\epsilon=0.1$.  With this in mind, the PN flux is the least performing approximation achieving the threshold error at separations of between $5\leq r/M\leq 10$ for both spins.  In each case the PN flux at 5.5PN performs exceptionally well, getting as close as $r=3.5M$ before having an error of $10^{-1}$.  However, this is clearly not reflective of convergence, as a 6PN flux could perform much worse.  The Pad\'e flux again improves over the PN flux, except at the 5.5PN order, reaching the threshold error at separations of between $3\leq r/M\leq 5$ in both cases, with the odd exception performance in the highest spin case at 5PN where a separation of $r=2.5M$ is reached.  However, the Pad\'e flux approximation now begins to suffer from singularities not only at 3.5PN order as we saw for lower spins, but also now at the 3PN order.  The 5.5PN inverted Chebyshev flux has a relative error of $10^{-1}$ at separations of around $r=5M$ in both cases.  While it achieves a consistent separation before reaching the threshold, it can only be marginally considered to be better than the PN flux, and is constantly outperformed by the Pad\'e flux.  However, the 5PN inverted Chebyshev flux reaches the threshold value at separations of approximately $r = 4M$ in both spin cases.  While sometimes outperformed at a particular approximation by either the PN or Pad\'e flux models, we believe that the 5PN inverted Chebyshev flux represents the most stable and consistently performing model for prograde orbits.

\section{Conclusion.}
\label{Conclusion}

In this work we extended the studies of~\cite{DIS,PS,Porter1} by taking into account the absorption of the GW by the central black hole, and by adding the information from the non-spinning case,  we created a 5.5PN Kerr flux. We saw that, because the 4PN flux is divergent, adding these higher order terms improves the convergence of the PN flux in all cases.

By first looking at the non-spinning case, we searched for the most convergent factorized flux at 5.5PN for each re-summation method. We tried a number of different combinations : the standard factorized flux given in Ref.~\cite{DIS} or the inverted factorized flux proposed in Ref.~\cite{PS}, and we also tried $x_{lso}$, the invariant speed at the last stable orbit and $x_{pr}$, the invariant speed at the photon ring, as a normalization factor in the logarithmic term of the flux. It appeared that for both Pad\'e and Chebyshev flux models, the best choice was to use the inverted flux normalized using $x_{lso}$, which confirms the choices made in~\cite{DIS} and~\cite{PS}, respectively. Then we compared the two resummation methods and the PN flux, at different orders, with a numerical flux.  We observed that, except when it presents a pole, the Pad\'e approximation is in general better than the PN flux.  However, while the standard Chebyshev flux performed extremely well, the better approximation was clearly the inverted Chebyshev flux function.

Next we studied the case of retrograde orbits, and once again we started by searching for the best factorized flux.   Contrary to the previous case, for negative spins the marginally better normalization choice is using $x_{pr}$, and as in the Schwarzschild case, the inverted flux is the best choice for the two techniques. When comparing the different fluxes we saw that, except for a few particular cases, the re-summation methods are better than the PN flux.  At times however, the PN flux has a fortunate crossing with the numerical flux near the LSO and therefore appears to work well.  However, the fluctuations is convergence from one order to another means that it is difficult to rely on the PN approximation.   In general, the Pad\'e approximation improves on the PN flux, but again there are orders of approximation where it performs worse than the PN approximation. The inverted Chebyshev approximation outperforms both other flux models, improving the accuracy of the flux model by a factor of two.

Our final study involved prograde orbits and once again we found that the best normalization is using $x_{lso}$, but while the inverted flux was the best choice for the Chebyshev approximation, it appeared that the standard flux is a better choice for the Pad\'e flux.  Here we found that the PN approximation performs worse and worse as we increase the spin magnitude, except at the 5.5PN order where it performs surprisingly well.   The Pad\'e flux is a great improvement over the PN flux, but again oscillates between good and bad performance.   In this case, we found that the inverted Chebyshev flux at 5.5PN order did not perform as well as expected.  Upon further investigation, we found that this was due to the factorized flux being divergent at this order.  However, by using the 5PN factorized flux, we were once again able to construct a very convergent Chebyshev flux model.  While more marginal, the inverted Chebyshev flux is the most stable model, converging to the same order of error at the LSO.

This work improves upon previous studies and introduces a new inverted Chebyshev flux model.  This model is highly convergent and stable.  In most cases, this flux model is more convergent at the 2.5PN level than either the PN or Pad\'e fluxes at 5.5PN order.  We believe that this model provides a good basis for the creation of theoretical templates for GW astronomy.

\appendix

\section{PN flux coefficients.}
\label{PN_coefficients}

Here we present the post-Newtonian coefficients taking into account the non-spinning terms up to $x^{11}$, the spinning terms up to $x^8$ and the absorption terms up to $x^{11}$.
\bea
a_{0} & = & 1 \\
a_{1} & = & 0 \\
a_{2} & = & -\frac{1247}{336} \\
a_{3} & = & 4\pi - \frac{11}{4}\,q \\
a_{4} & = & -\frac{44711}{9072} + \frac{33}{16}\,q^{2} \\
a_{5} & = & -\frac{8191\,\pi}{672} - \frac{63}{16}\,q- \frac{3}{4}\,q^3\\ 
a_{6} & = & \frac{6643739519}{69854400}-\frac{1712\,\gamma}{105}+\frac{16\,\pi^{2}}{3}-\frac{3424\,\ln 2}{105} - \frac{65\,\pi}{6}q \\ 
a_{7} & = & -\frac{16285\,\pi}{504}
  + \frac{158147}{3888}\,q
  + \frac{65\,\pi}{8}\,q^{2}
  - \frac{241}{48}\,q^{3}\\ 
a_{8} & = & -\frac{321516361867}{3178375200}
   + \frac{232597\,\gamma}{4410} 
   - \frac{1369\,\pi^{2}}{126}
   + \frac{39931\,\ln 2}{294}
   - \frac{47385\,\ln 3}{1568}
   + \frac{1}{2}\,\kappa \nonumber \\
& - & \frac{359\,\pi}{14}\,q
   + \frac{49127}{4536}\,q^2
   + \frac{13}{2}\,\kappa\,q^2
   + \frac{13}{16}\,q^4
   + 3\,q^4\kappa   
   + 2\,q\,B_2
   + 6\,q^3{B_2} \\
a_{9} & = & \frac{265978667519\,\pi}{745113600} 
   - \frac{6848\,\gamma\,\pi}{105} 
   - \frac{13696\,\pi\,\ln 2}{105}
   - \frac{43}{7}\,q
   - \frac{4651}{336}\,q^3
   - \frac{17}{56}\,q^5 \\
a_{10} & = & -\frac{2495197796217283}{2831932303200} 
  + \frac{916628467\,\gamma}{7858620} 
  - \frac{424223\,\pi^2}{6804} 
  - \frac{83217611\,\ln 2}{1122660} 
  + \frac{47385\,\ln 3}{196}
  + 2\,\kappa  \nonumber \\
& + & \frac{163}{8}\,\kappa\,q^2
  + \frac{433}{24}\,q^2
  + \frac{33}{4}\,q^4\kappa
  - \frac{95}{24}\,q^4
  + qB_1
  - \frac{3}{4}\,q^3B_1
  + 6\,qB_2
  + 18\,q^3B_2\\
a_{11} & = & \frac{8399309750401\,\pi }{101708006400} 
  + \frac{177293\,\gamma\,\pi }{1176}
  + \frac{8521283\,\pi \,\ln 2}{17640} 
  - \frac{142155\,\pi \,\ln 3}{784} \nonumber \\
& + & \frac{428}{105}\,\gamma\,q
  + \frac{428}{105}\,q\ln\kappa
  - \frac{2586329}{44100}\,q
  + \frac{2}{3}\,\pi^2q
  - 31\,\frac {q}{\kappa}
  + \frac{7}{3}\,\kappa\,q
  + \frac{428}{105}\ln2\,q \nonumber \\
& + & \frac{428}{35}\,q^3\gamma
   - 32\,\frac {q^3}{\kappa}
  + \frac{428}{35}\,q^3\ln2
  + 2\,q^3\pi^2
  + \frac{428}{35}\,q^3\ln\kappa
  + \frac{227}{6}\,\kappa\,q^3
  - \frac{1640747}{19600}\,q^3 \nonumber \\
& + & 57\,\frac{q^5}{\kappa}
  + 19\,q^5\kappa
  + \frac{455}{16}\,q^5
  + 6\,\frac{q^7}{\kappa}
  + \frac{428}{105}\,q{A_2}
  + \frac{428}{35}\,q^3{A_2}
  - \frac{4}{3}\,q^2{B_1}
  + q^4{B_1} \nonumber \\
& - & 4\,B_2
   - 4\,\frac{B_2}{\kappa}
   - 48\,\frac{q^2{B_2}}{\kappa}
   - 44\,q^2B_2
   + 28\,\frac{q^4{B_2}}{\kappa}
   + 24\,\frac{q^6{B_2}}{\kappa}
   - 8\,qB_2^2
   - 24\,q^3B_2^2 \nonumber \\
& - & 8\,\frac {q{C_2}}{\kappa}
  - 4\,q\,C_2
  - 24\,\frac{q^3{C_2}}{\kappa}
  - 12\,q^3C_2 \nonumber \\
b_{6} & = & -\frac{1712}{105}\\
b_{7} & = & 0 \\
b_{8} & = & \frac{232597}{4410} \\
b_{9} & = & - \frac{6848\,\pi}{105} \\
b_{10} & = & \frac{916628467}{7858620} \\
b_{11} & = & \frac{177293\,\pi}{1176} + \frac{856}{105}\,q + \frac{856}{35}\,q^3
\eea
where $\gamma$ is the Euler's number, $\kappa=\sqrt{1-q^2}$, and the $A_n$, $B_n$, and $C_n$ are defined by
\begin{eqnarray}
A_n&=&{1\over 2}\left[\psi^{(0)}\left(3+{n i q\over \sqrt{1-q^2}}\right)
+\psi^{(0)}\left(3-{n i q\over \sqrt{1-q^2}}\right)
\right], \cr \nonumber\\ \nonumber \\ \nonumber
B_n&=&{1\over 2i}\left[\psi^{(0)}\left(3+{n i q\over \sqrt{1-q^2}}\right)
-\psi^{(0)}\left(3-{n i q\over \sqrt{1-q^2}}\right)
\right], \cr \\ \nonumber \\ 
C_n&=&{1\over 2}\left[\psi^{(1)}\left(3+{n i q\over \sqrt{1-q^2}}\right)
+\psi^{(1)}\left(3-{n i q\over \sqrt{1-q^2}}\right)
\right],
\end{eqnarray}
where $\psi^{(0)}(z)$ and $\psi^{(1)}(z)$ are respectively the digamma and the trigamma functions.  For ease and speed of calculation,  the digamma and trigamma functions were generated using a Chebyshev-Pad\'e approximation such that the maximum error between the approximation and the analytic expressions is less than $10^{-10}$ for all spin values.


\begin{thebibliography}{}

\bibitem{AABBB1}
Amaro-Seoane~P \emph{et al.}, arXiv:1201.3621 [astro-ph.CO]

\bibitem{AABBB2}
Amaro-Seoane~P \emph{et al.}, arXiv:1202.0839 [gr-qc]

\bibitem{ET1}
Punturo~M et. al., \emph{Class.~Quantum Grav.}~{\bf 27} 194002 (2010)

\bibitem{ET2}
Sathyaprakash~B \emph{et. al.}, \emph{Class.~Quantum Grav.}~{\bf 29} 124013 (2012)

\bibitem{Poisson2} 
Poisson~E, \emph{Phys. Rev. D} {\bf 49}, 1860 (1993)

\bibitem{SSTT} 
Shibata~M, Sasaki~M, Tagoshi~H and Tanaka~T, \emph{Phys. Rev. D} {\bf 51}, 1646 (1995) 

\bibitem{TSTS} 
Tagoshi~H, Shibata~M, Tanaka~T and Sasaki~M, \emph{Phys. Rev. D} {\bf 54}, 1439 (1996)

\bibitem{Poisson1}
Poisson~E, \emph{Phys. Rev. D} {\bf 47}, 1497 (1993)

\bibitem{Cutetal1}
Cutler~C, Finn~L~S, Poisson~E and Sussman~G~J, \emph{Phys. Rev. D} {\bf 47}, 1511 (1993)

\bibitem{TagNak} 
Tagoshi~H and Nakamura~T, \emph{Phys. Rev. D} {\bf 49}, 4016 (1994)

\bibitem{TagSas} 
Tagoshi~H and Sasaki~M, \emph{Prog. Theor. Phys.} {\bf 92}, 745 (1994)

\bibitem{TTS} 
Tanaka~T, Tagoshi~H and Sasaki~M, \emph{Prog. Theor. Phys.}, 1087 (1996)

\bibitem{Brady}
Brady~P, Creighton~J.D.E. and Thorne~K.S.,  \emph{Phys. Rev. D} {\bf 58}, 061501 (1998)

\bibitem{YB}
Yunes~N and Berti~E, \emph{Phys. Rev. D} {\bf 77}, 124006 (2008)

\bibitem{ZYB}
Zhang~Z, Yunes~N and Berti~E, \emph{Phys. Rev. D} {\bf 84}, 024029 (2011)

\bibitem{DIS}
Damour~T, Iyer~B~R and Sathyaprakash~B~S, \emph{Phys. Rev. D}, {\bf 57} 885 (1998)

\bibitem{PS}
Porter~E~K and Sathyaprakash~B~S, \emph{Phys. Rev. D}, {\bf 71} 024017 (2005)

\bibitem{Porter1}
Porter~E~K, \emph{Class.~Quantum Grav.} {\bf 23}, S837 (2006)

\bibitem{Porter2}
Porter~E~K, \emph{Phys. Rev. D} {\bf 76}, 104002 (2007)

\bibitem{Helstrom}
Helstrom~C~W, \emph{Statistical Theory of Signal Detection} (Pergamon Press, London, 1968)

\bibitem{BPT}
Bardeen~J~N, Press~W~H and Teukolsky~S~A, \emph{Astrophys. J.} {\bf 178}, 374 (1972)

\bibitem{Chand}
Chandrasekhar~S, \emph{The Mathematical Theory of Black Holes} (Oxford Uni. Press, New York, 1983)

\bibitem{MSSTT}
Mino~Y, Sasaki~M, Shibata~M, Tagoshi~H and Tanaka~T, \emph{Prog. Theor. Phys. Suppl}, 128, 1 (1997)

\bibitem{BO}
Bender~C~M and Orszag~S~A, \emph{Advanced Mathematical Methods for Scientists and Engineers}, McGraw-Hill International Book Company, Singapore, 1978

\bibitem{CP}
Carr\'e~J and Porter~E~K, arXiv:1112.3222v1 [gr-qc]

\bibitem{MH}
Mason~J~C and Handscomb~D~C, \emph{Chebyshev Polynomials} (Chapman and Hall/CRC, Boca Raton, 2003)

\bibitem{NR}
Press~W~H, Flannery~B~P, Teukolsky~S~A and Vetterling~W~T, \emph{Numerical recipes in C~: The Art of Scientific Computing} (Cambridge Uni. Press, Cambridge 1992) 2nd Edition




\end{thebibliography}
\end{document}